\documentclass[12pt]{iopart}
\usepackage{graphicx}
\usepackage{color}
\begin{document}

\title[Magnetic properties of the Ag-In-RE approximants]{Magnetic properties of the Ag-In-rare-earth 1/1 approximants}

\author{Soshi~Ibuka, Kazuki~Iida\footnote{Present address: Department of Physics, University of Virginia, Charlottesville, Virginia 22904-4714, USA}, and Taku~J~Sato}

\address{Neutron Science Laboratory, Institute of Solid State Physics,
University of Tokyo, 106-1 Shirakata, Tokai, Ibaraki 319-1106, Japan}
\ead{ibuka@issp.u-tokyo.ac.jp}

\begin{abstract}
We have performed magnetic susceptibility and neutron scattering measurements on polycrystalline Ag-In-RE (RE: rare-earth) 1/1 approximants. In the magnetic susceptibility measurements, for most of the RE elements, inverse susceptibility shows linear behaviour in a wide temperature range, confirming well localized isotropic moments for the RE$^{3+}$ ions.
Exceptionally for the light RE elements, such as Ce and Pr, non-linear behaviour was observed, possibly due to significant crystalline field splitting or valence fluctuation.
For RE = Tb, the susceptibility measurement clearly shows a bifurcation of the field-cooled and zero-field-cooled susceptibility at $T_{\rm f} = 3.7$~K, suggesting a spin-glass-like freezing.
On the other hand, neutron scattering measurements detect significant development of short-range antiferromagnetic spin correlations in elastic channel, which accompanied by a broad peak at $\hbar\omega = 4$~meV in inelastic scattering spectrum.
These features have striking similarity to those in the Zn-Mg-Tb quasicrystals, suggesting that the short-range spin freezing behaviour is due to local high symmetry clusters commonly seen in both the systems.
\end{abstract}


\submitto{\JPCM}

\maketitle

\section{Introduction}
Static and dynamic behaviour of spins in quasiperiodic structure has been of considerable interest from early days of the quasicrystal research to date~\cite{sato05}.
Theoretically, several quasiperiodic lattices with localized moments have been studied in details, and a number of intriguing magnetic orderings, including long-range quasiperiodic order, have been proposed to date~\cite{ved04,wes03}.
Experimentally, however, model materials which contain localized magnetic moments are quite limited even today, and thus nature of real magnetic quasicrystals remain largely unaddressed.
One of the rare examples of magnetic quasicrystals is the Zn-Mg-RE (RE: rare-earth elements) icosahedral quasicrystal~\cite{luo93,nii94}.
For this material, existence of localized RE$^{3+}$ ions was confirmed by the magnetic susceptibility, which follows the Curie-Weiss law in a wide temperature range~\cite{hat95,fis99}.
Below a certain low temperature $T_{\rm f} = 5.8$~K, a bifurcation of field-cooled (FC) and zero-field-cooled (ZFC) susceptibility was clearly seen. At first sight, this seems that spins in the Zn-Mg-RE quasicrystal freeze randomly, as in canonical spin-glasses.
However, neutron scattering results revealed that short-range-spin correlations develop with collective localized excitations as temperature decrease~\cite{sato06}, which striking contrast to the absence of spin correlations in canonical spin glasses~\cite{mur81}.
From these observations, it is now established that, on cooling, antiferromagnetic spin correlations first develop in high-symmetry dodecahedral clusters of RE ions, which are essentially present in Zn-Mg-RE icosahedral quasicrystals.
Spins in each cluster fluctuate coherently with no inter-cluster correlation.
On further cooling to the temperature $T_{\rm f}$, cluster-spin fluctuations freeze in a cluster-spin-glass state without inter-cluster long-range magnetic ordering.

Then, a natural question arises. Why does no long-range magnetic order occurs at the low temperature and why the magnetic correlations are limited to short-range order?
Two reasons might be responsible for this: one is that the quasiperiodicity of the quasicrystal limits the spatial extension of spin correlations. The other is that the spin correlations develop only in the intrinsic high-symmetry RE3+ spin clusters, and correlations between the clusters are interrupted with chemical or structural disorder which commonly exists in quasicrystals and approximants.

To draw a conclusion on this issue, it is worthwhile to study magnetic ordering behaviour and spin correlations in magnetic approximants, which have a periodic lattice and include periodically arranged magnetic clusters which consist of localized spins, contrary to quasicrystals.

After the discovery of an icosahedral quasicrystal in ${\rm Ag_{42}In_{42}Yb_{16}}$ by Guo and Tsai~\cite{Guo2002}, the formation of the 1/1 approximants were identified in the most of Ag-In-RE alloys~\cite{Rua04, mor08}.
These 1/1 approximants are considered as isostructural to the well-known Cd$_6$Yb approximant~\cite{Joh1964, Pal1971, Tak2001, Gom2003}, which is a bcc crystal of the space group {\it Im$\bar{3}$} with icosahedral clusters of RE elements and forms the same local structure as the stable binary icosahedral quasicrystal Cd$_{5.7}$Yb~\cite{Guo2000, Tsa2000, takakura07}.
This is the first discovery of the approximant phase which contains magnetic rare-earth elements and doesn't contain neutron-absorbing elements such as Cd.
Therefore these approximants enable us to perform a microscopic-neutron-scattering investigation on magnetic properties of high-symmetry magnetic clusters arranged periodically.
Although the local atomic structure is different from that in the Zn-Mg-RE quasicrystals~\cite{ishimasa98,ishimasa04}, in which the symmetry of the RE clusters would be dodecahedral, to study the magnetic properties of the Ag-In-RE approximant will provide a key to solve the previous question in view of high symmetry of the icosahedral magnetic clusters.
In this work, we have performed magnetic susceptibility and neutron scattering measurements on polycrystalline Ag-In-RE 1/1 approximants.
The magnetic susceptibility versus temperature for twelve Ag-In-RE (RE = Ce, Pr, Nd, Sm, Eu, Gd, Tb, Dy, Ho, Er, Tm and Yb) showed no anomaly owing to long-range magnetic order.
For most of the samples, the inverse susceptibility follows linear behaviour above 100~K, which suggests RE ions in these samples are isotropically well localized.
At low temperature near 3~K, a spin-glass-like bifurcation of the FC and ZFC magnetic susceptibility was observed for RE = Eu, Gd, Tb and Dy.
In the neutron elastic and inelastic scattering measurements, which were performed for RE = Tb, development of diffuse scattering was observed as temperature decrease in addition to a broad peak at $\hbar\omega=4$~meV. These results indicate that the spin fluctuations freeze into cluster-spin-glass with short-range magnetic order, which highly correspond to those in the Zn-Mg-Tb quasicrystal.

\section{Experimental details}
For magnetic susceptibility measurements, polycrystalline samples of the Ag-In-RE (RE = Ce, Pr, Nd, Sm, Eu, Gd, Tb, Dy, Ho, Er, Tm and Yb) 1/1 approximants were prepared by means of melting constituent elements in an arc furnace.
Purity of the starting elements were 99.9999\% for Ag, 99.9999\% for In and 99.9\% for RE.
Most of nominal compositions for the polycrystal preparation were decided by the earlier report~\cite{mor08} and are shown in figure~1.
The resulting as-cast alloys were wrapped into Mo foils, sealed in quartz tubes under pure Ar atmosphere and subsequently heat-treated at 823~K for 100~h.
For the neutron scattering experiments, different procedure was chose to prepare as much as 10 gs of an alloy; A polycrystalline sample of Ag$_{49}$In$_{37}$Tb$_{14}$ was synthesized directly from the constituent elements in a electric furnace. The elements were inserted in a high-purity Al$_2$O$_3$ crucible, sealed in a quartz tube, heat-treated at 1273~K for 10~h and again heat-treated at 923~K for 100~h.

Phase quality of the resulting samples was characterized by X-ray powder diffraction (Rigaku, Miniflex) and scanning electron microscopy (SEM) (JEOL, JSM-5600).
In addition, their compositions were determined by energy dispersive X-ray spectroscopy (Oxford, Link ISIS).
Magnetic susceptibility was measured between 2 and 300~K by a superconducting quantum interference device (SQUID) magnetometer (Quantum Design, MPMS-XL) with an external d.c. field of 100~Oe.
Neutron elastic, quasielastic and inelastic scattering experiments were carried out on the poly crystalline Ag$_{49}$In$_{37}$Tb$_{14}$ sample.
For thermal neutron experiments, the triple-axis spectrometer ISSP-GPTAS was used, which installed at the JRR-3 research reactor (Tokai, Japan).
To minimize neutron absorption effect of In atom, the powdered sample was inserted in a double-cylindrical annular Al cell with sample thickness of $t = 1.5$~mm.
For the elastic scattering experiment, the spectrometer was operated in a double-axis mode with incident energy of $E_{\rm i} = 13.7$~meV and with collimations of 40'-80'-40'.
To eliminate higher harmonic neutrons, two pyrolytic-graphite (PG) filters were inserted between the source and monochrometer, and between the sample and analyzer.
For the thermal-neutron inelastic scattering experiment, a triple-axis mode was employed with doubly (horizontally and vertically) focusing analyzer selecting finial neutrons with $E_{\rm f} = 14.7$~meV.
The incident side collimations were 40'-40', whereas a radial collimator (Rad) and a slit of 30~mm width were inserted in the outgoing path. 
The energy resolution was 1.3~meV in full width at half maximum (FWHM) at the elastic position.
Furthermore, for quasielastic scattering experiments in higher energy-resolution, the cold-neutron triple-axis spectrometer ISSP-HER were used, which installed at the C1 guide tube of JRR-3.
Doubly focusing analyzer was employed to fix the final energy to $E_{\rm f} = 3.0$~meV with the collimations Guide-Open-Rad-Slit(20~mm), yielding an energy resolution of 0.1~meV (FWHM) at the elastic position.
Finally, powder neutron inelastic scattering experiments on Zn-Mg-Tb magnetic quasicrystal were performed as an supplement with \cite{sato06}, using the inverted-geometry time-of-flight spectrometers LAM-40~\cite{lam40}, installed at KENS, KEK, Japan.
Pyrolytic graphite (PG) 002 reflections were used for the analyzer to fix the final energy at $E_{\rm f} = 4.9$~meV, and the energy resolution was estimated using a vanadium standard as $0.32$~meV (FWHM) at the elastic position.

\section{Sample characterization}
\begin{figure}[htbp]
\includegraphics[angle=-90,width=120mm]{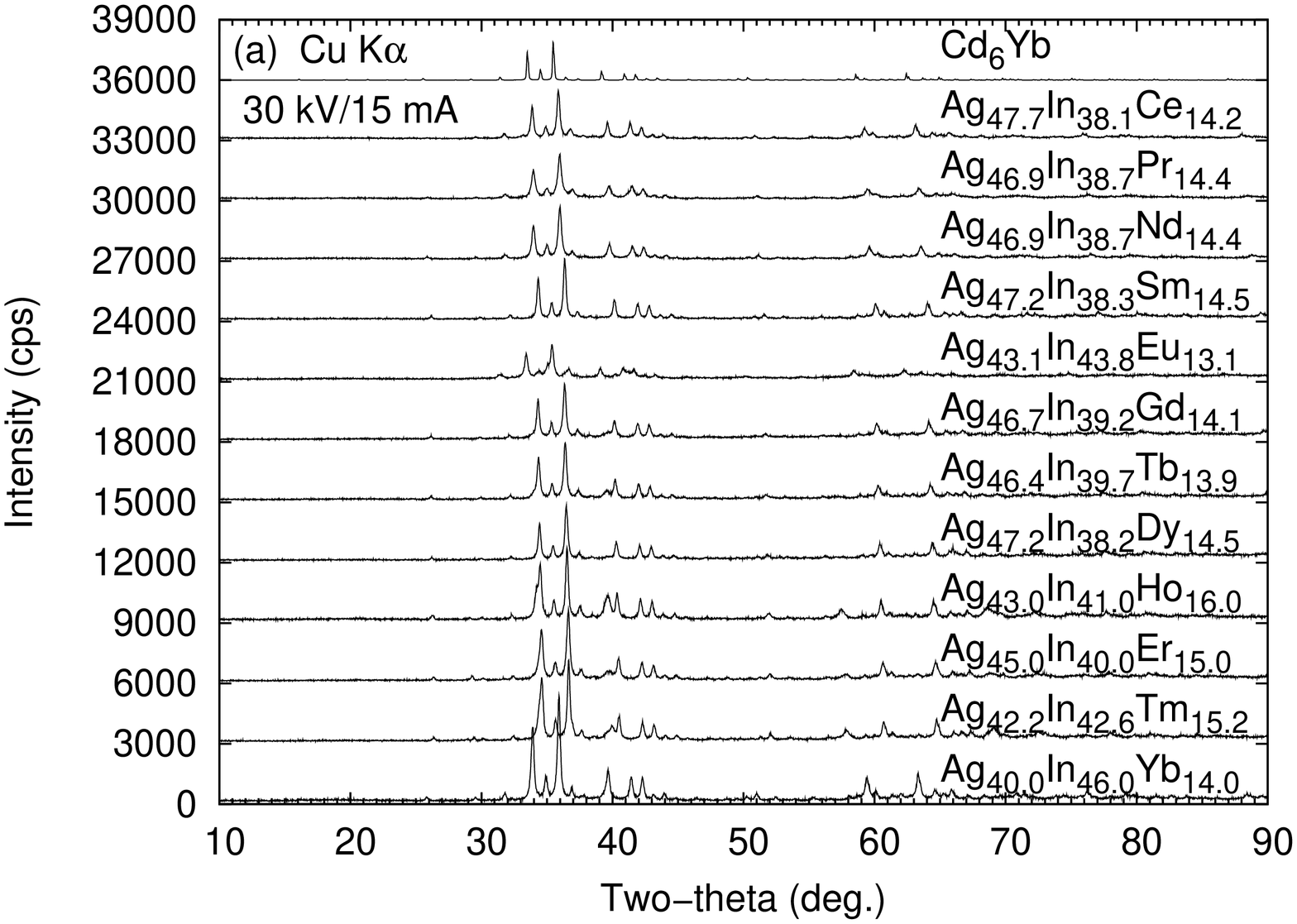}
\includegraphics[width=80mm]{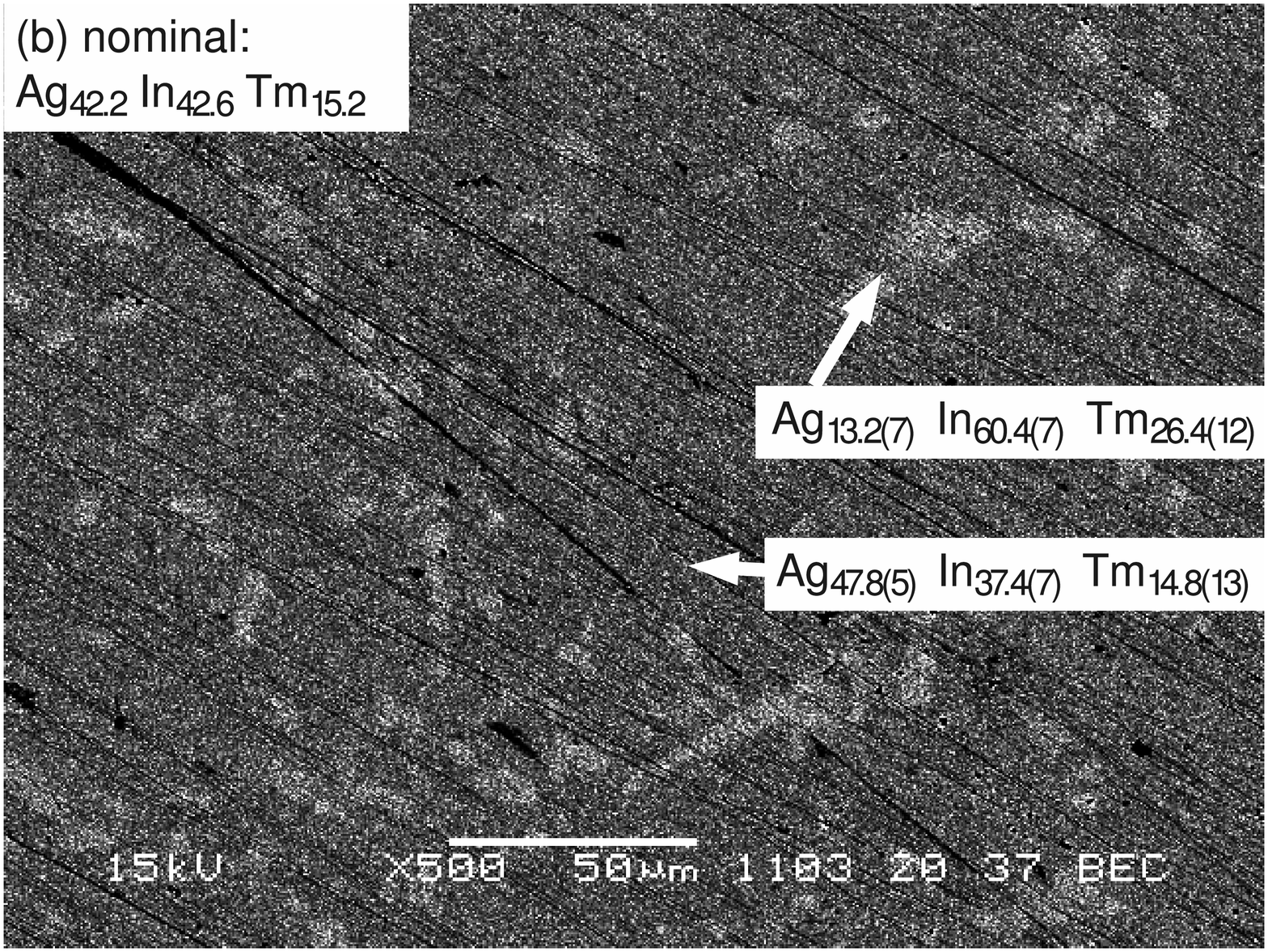}
\includegraphics[width=80mm]{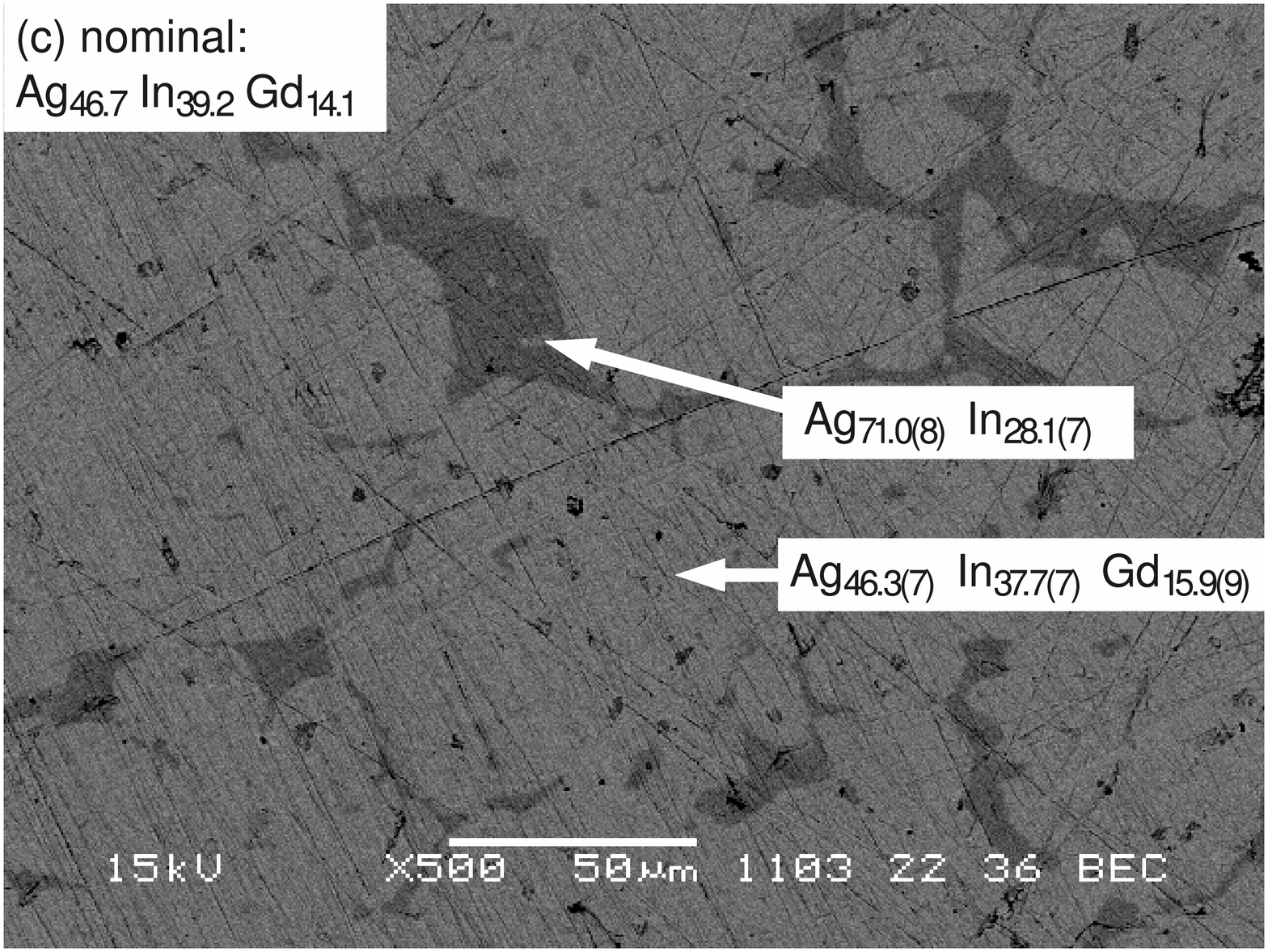}
\caption{(a) X-ray powder diffraction patterns of the prepared Ag-In-RE samples. Calculated powder diffraction pattern for Cd$_6$Yb 1/1 approximant phase is also given for comparison~\cite{Tak2001}. Nominal compositions are shown on the right side.
(b)(c) SEM micrographs of the Ag-In-Tm and Ag-In-Gd sample respectively, which were taken with incident electron energy of 15~kV.
}
\label{xraybec}
\end{figure}

Figure~1(a) shows the X-ray powder diffraction patterns of the prepared polycrystalline samples.
As seen in the figure, formation of the 1/1 approximant phase was confirmed as major phase for all the RE elements.
On the other hand, small impurity peaks were also found for RE = Gd, Tb, Ho, Er and Tm, which indicate small amount of contaminating phases exist.
Representative examples of SEM micrographs for RE = Tm and Gd are shown in figure~1(b) and 1(c).
In each micrograph, we observed two phases with different compositions; composition of the major phase was roughly equal to the nominal one, which also suggests the main phase was 1/1 approximant. The other minority phase was Ag$_3$In or Ag$_{10}$In$_{65}$RE$_{25}$.
Such contamination was more or less observed in every RE sample, including the Ag-In-Tb sample for neutron experiments.
Each impurity phase was, however, tiny in volume, thus the deviation of the nominal composition from the ideal one of the 1/1 approximant phase was sufficiently small.
From these results, we can conclude that 1/1 approximants were formed as the major phase in all the Ag-In-RE samples.

\section{Magnetic susceptibility}
\begin{figure}[htbp]
\includegraphics[angle=-90,width=120mm]{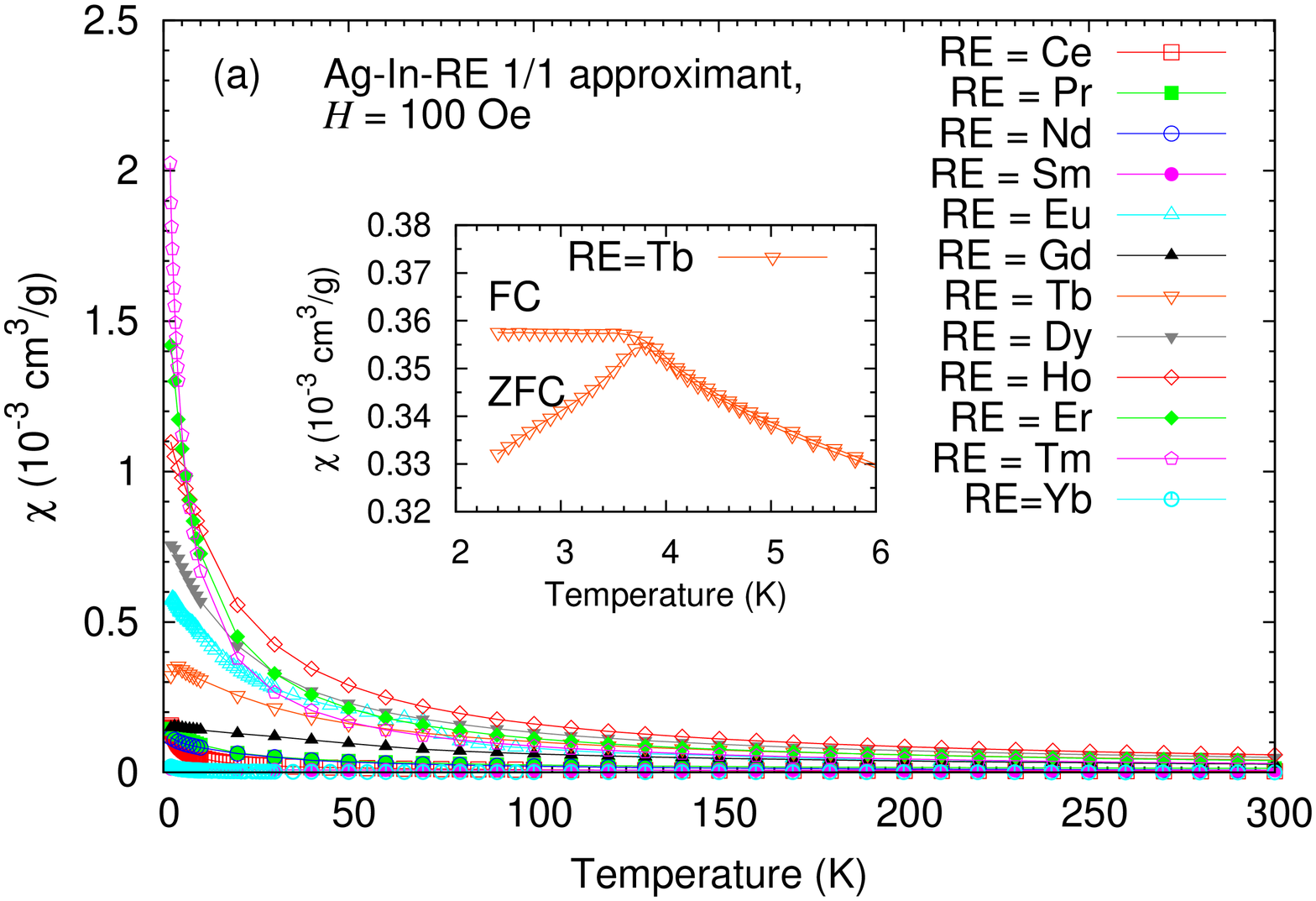}

\includegraphics[angle=-90, width=120mm]{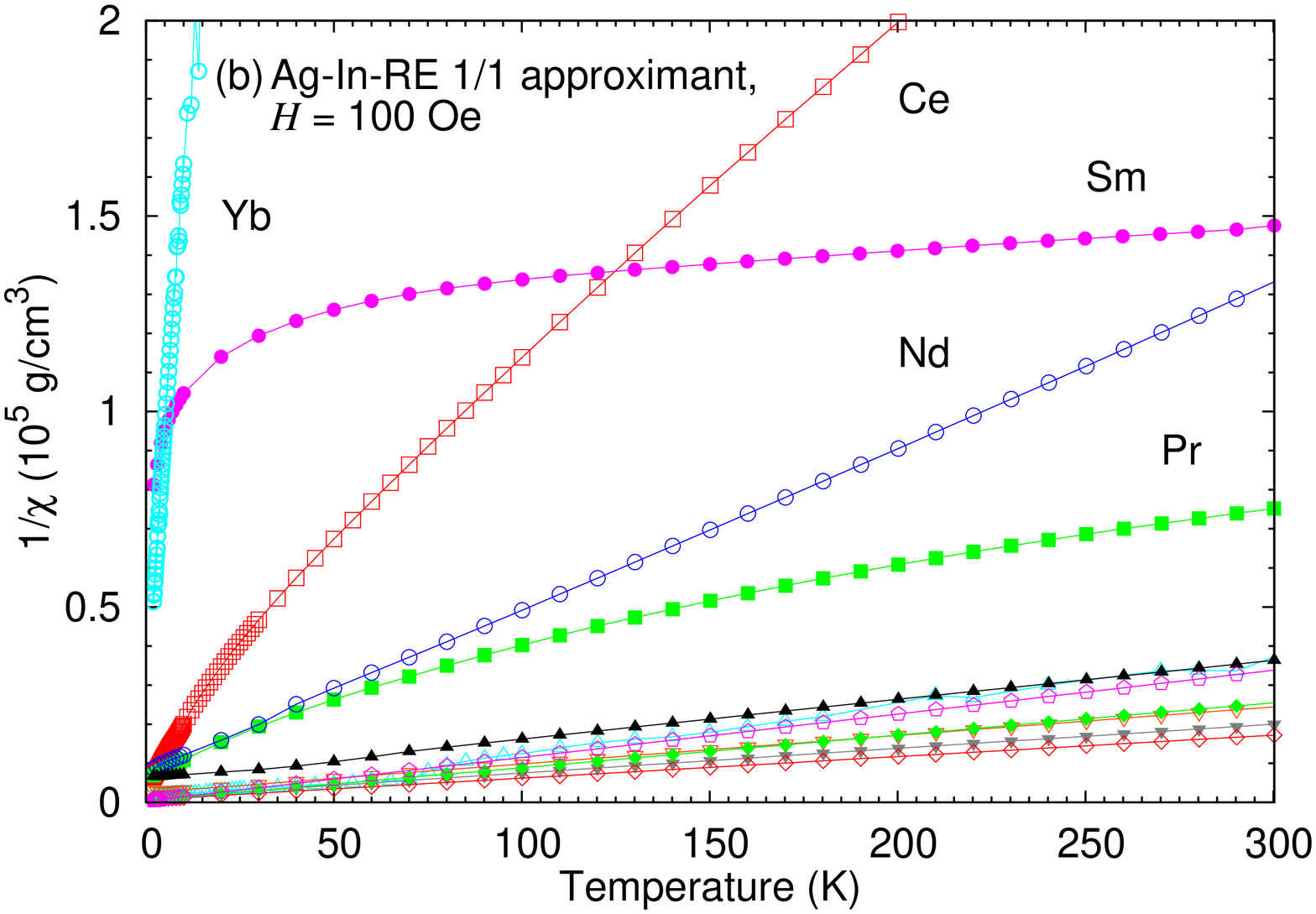}
\caption{(a) Temperature dependence of magnetic susceptibility for all the RE samples prepared in the present work observed under the external field $H_{\rm ext} = 100$~Oe. 
Inset: field-cooled and zero-field-cooled susceptibility for RE = Tb in the low temperature region.
(b) Inverse susceptibility for all the RE samples.}
\label{susceptibility}
\end{figure}

\begin{table}[htbp]
\caption{\label{mageff}Observed Weiss temperatures $\theta$, effective moment sizes $\mu_{\rm eff}$, freezing temperatures $T_{\rm f}$ and calculated moment sizes for free RE$^{3+}$ (and for RE$^{2+}$ for RE = Eu).}
\lineup
\begin{indented}
\item[]\begin{tabular}{@{}lllll}
\br
RE & $\mu_{\rm eff}$ & $\mu_{\rm RE^{+3}}~{\rm calc.}$ & $\theta({\rm K})$ & $T_{\rm f}({\rm K})$\\
\mr
Nd& \04.543(4)& \03.6              & \-23.23(18) & ---\\
Eu& \08.2(8)  & \00.0(7.94$^{\rm a}$)& \06(16)     & 2.5\\ 
Gd& \08.89(3) & \07.94             & \-55.5(9)   & 3.3\\
Tb& 10.77(7)  & \09.72             & \-34.13(14) & 3.7\\
Dy& 11.262(10)& 10.6               & \-17.69(18) & 2.5\\
Ho& 11.638(7) & 10.6               & \-12.09(12) & ---\\
Er& \09.746(9)& \09.59             & \0\-5.58(18)& ---\\
Tm& \08.490(4)& \07.57             & \0\-3.96(6) & ---\\
\br
\end{tabular}
\item[] $^{\rm a}$ for ${\rm RE^{+2}}$.
\end{indented}
\end{table}

Temperature dependence of magnetic susceptibility for all the prepared polycrystalline samples is shown in figure~2(a).
No clear anomaly due to long-range magnetic order could be observed for all the RE samples.
We note that a weak anomaly for RE = Eu is due to ferromagnetic transition of Eu$_2$O$_3$ impurity phase, which inevitably appears in the polycrystalline grains because of highly oxidizing nature of Eu.
For RE = Sm and Yb, the magnetic susceptibility is very small, showing only impurity upturn at the low temperatures.
This suggests that these RE ions are in the non-magnetic divalent state.
Inverse susceptibility is also shown in figure~2(b).
The inverse susceptibility for RE = Ce and Pr is not linear even at the room temperature, indicating that there is considerable crystal-field-splitting corresponding to the energy scale of 300~K or valence fluctuation effect~\cite{Var76}.
In the other RE samples (RE = Nd, Eu, Gd, Tb, Dy, Ho, Er and Tm), the inverse susceptibility shows linear behaviour above 100~K.
An effective moment size $\mu_{\rm eff}$ and a Weiss temperature $\theta$ were obtained by fitting the inverse susceptibility to a Curie-Weiss law:
\begin{equation}
\frac{1}{\chi} = \left[ \frac{N\mu_{\rm eff}^2\mu_{\rm B}^2}{3 k_{\rm B} (T - \theta)} + \chi_0 \right]^{-1},
\end{equation}
where $k_{\rm B}$, $N$, $\mu_{\rm B}$ and $\chi_0$ are the Boltzmann factor, the number of magnetic ions in unit volume, the Bohr magneton and temperature independent susceptibility, respectively.
$\mu_{\rm eff}$ and $\theta$ are given in table~1 together with calculated magnetic moment sizes for free RE$^{3+}$ (and RE$^{2+}$ for RE = Eu) ions.
As can be seen in the table, observed effective moments are in good agreement with theoretical free ion values for RE$^{3+}$ (or RE$^{2+}$ for RE = Eu), confirming well localized isotropic nature of the magnetic moments.
Except for RE = Eu, $\theta$ is negative, indicating dominant antiferromagnetic interactions between magnetic moments.
It is noteworthy that $\theta$ decreases as the RE becomes heavier from Gd to Tm, namely $\theta$ is well scaled by de Gennes factor, suggesting electronic states are equal at Fermi level for different RE samples~(cf.~\cite{hat95}).

At low temperature of $\sim 3$~K, a spin-glass-like bifurcation of the FC and ZFC magnetic susceptibility was observed for RE = Eu, Gd, Tb and Dy.
Representative susceptibility at low temperature for RE = Tb is shown in the inset of figure~2(a).
A clear bifurcation is seen at and below the freezing temperature $T_{\rm f} = 3.7$~K, suggesting a spin-glass-like freezing.
The temperature $T_{\rm f}$ is also listed in table~1.
We note that for RE = Tb, $T_{\rm f}$ is the highest among those which show the freezing behaviour.
A ratio $|\theta/T_{\rm f}|$ is approximately 10 for RE = Tb, and thus this is classified as a highly frustrated magnet.

\section{Neutron scattering}
\begin{figure}[htbp]
\includegraphics[width=80mm]{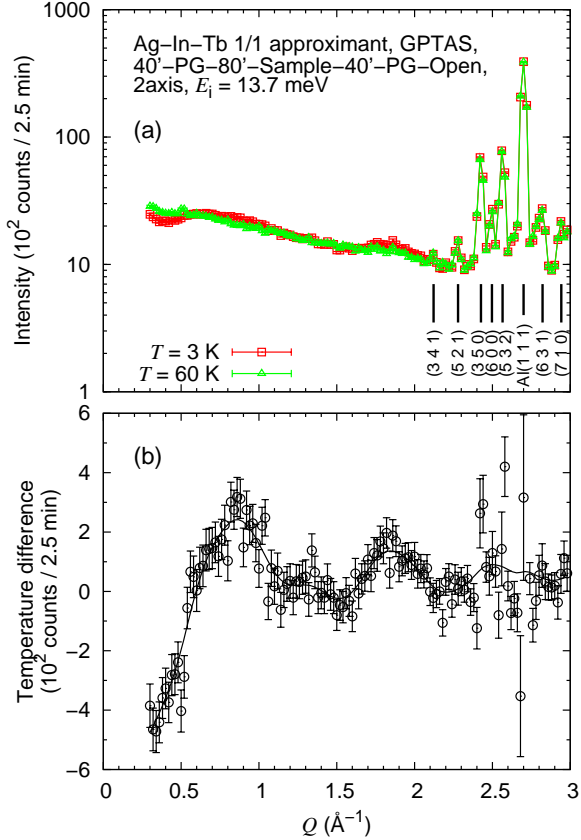}
\caption{(a) Neutron powder diffraction patterns at $T = 60$~K (open triangles) and 3~K (open squares) in Ag-In-Tb 1/1 approximant observed in double-axis mode.
Vertical solid lines at the bottom represent nuclear Bragg positions for the 1/1 approximant phase.
(b) The open circles stand for the difference between $T = 3$~K and $60$~K patterns~($I(3~{\rm K})-I(60~{\rm K})$). The solid line is a guide to the eye. }
\label{4Gelasticneutron}
\end{figure}

\begin{figure}[htbp]
\includegraphics[angle=-90, width=80mm]{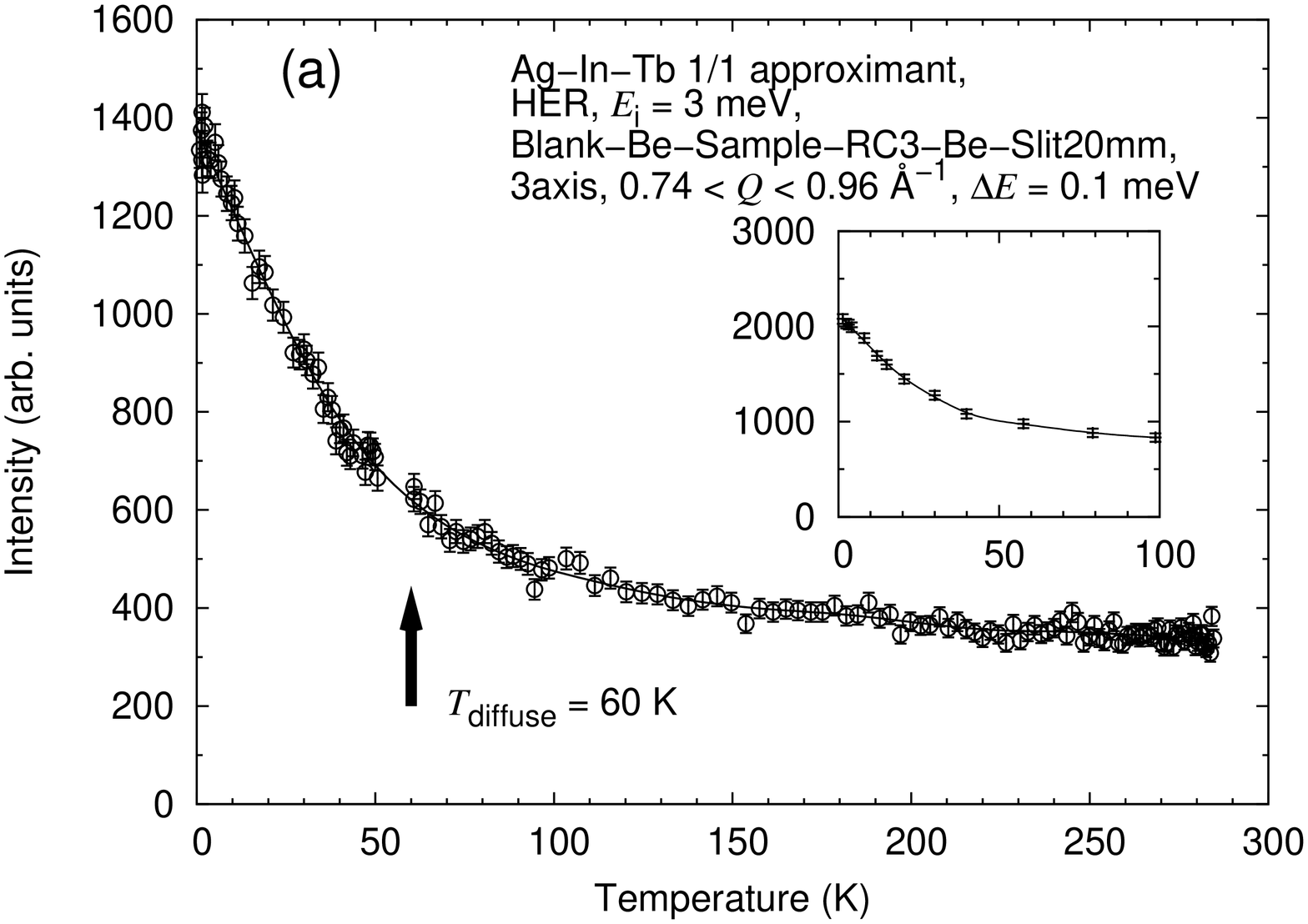}
\includegraphics[angle=-90, width=80mm]{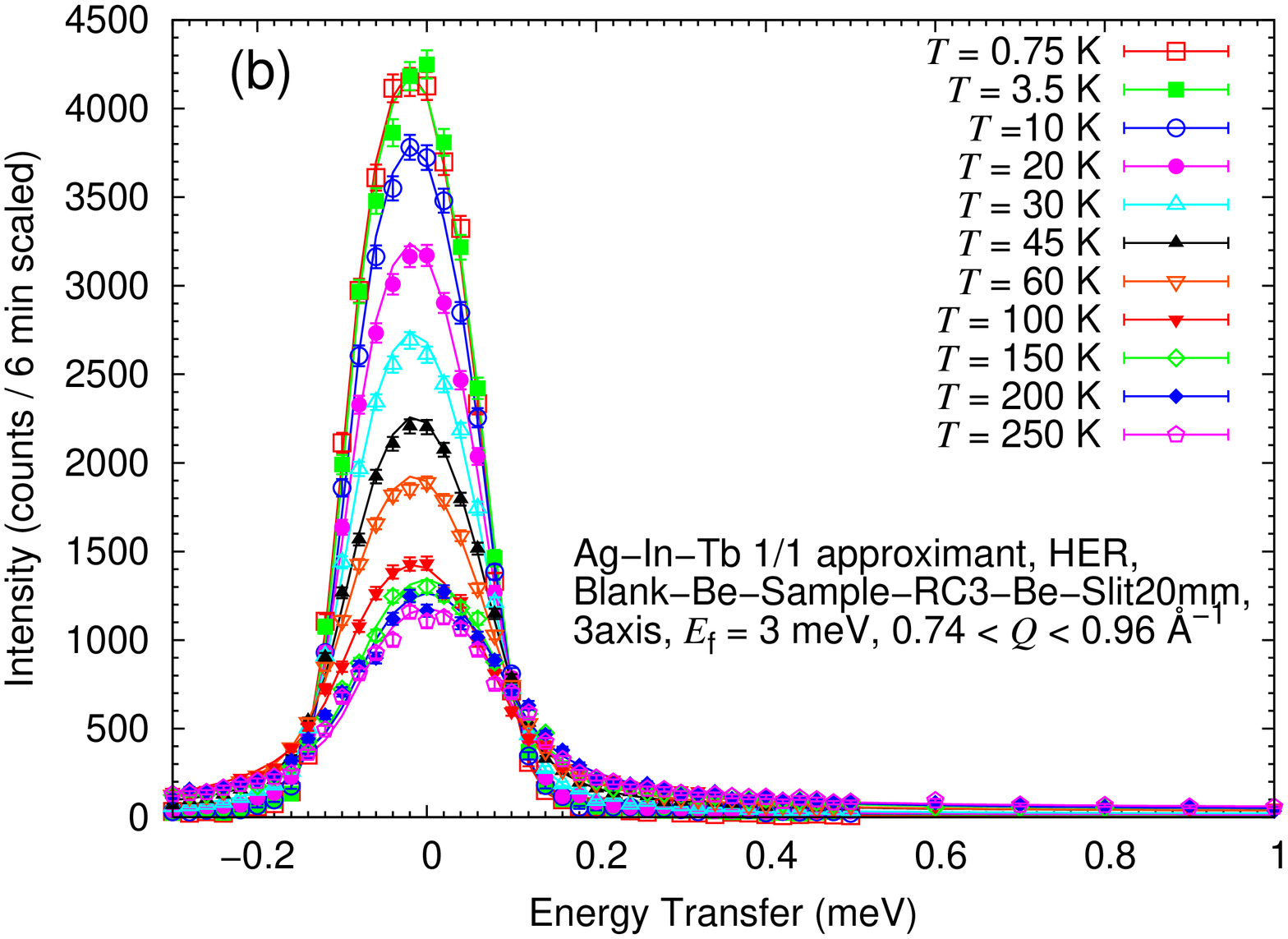}\\
\includegraphics[angle=-90, width=80mm]{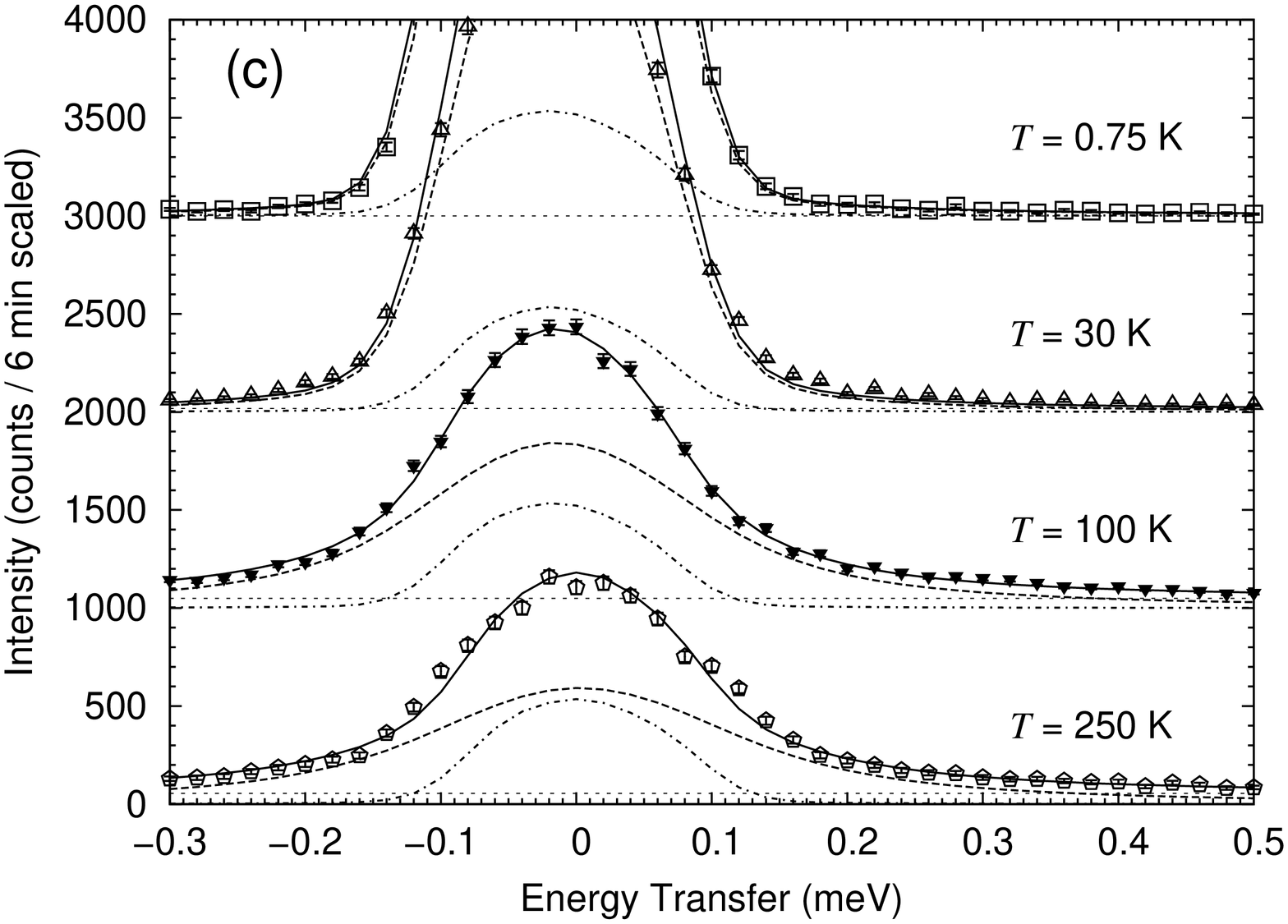}
\includegraphics[angle=-90, width=70mm]{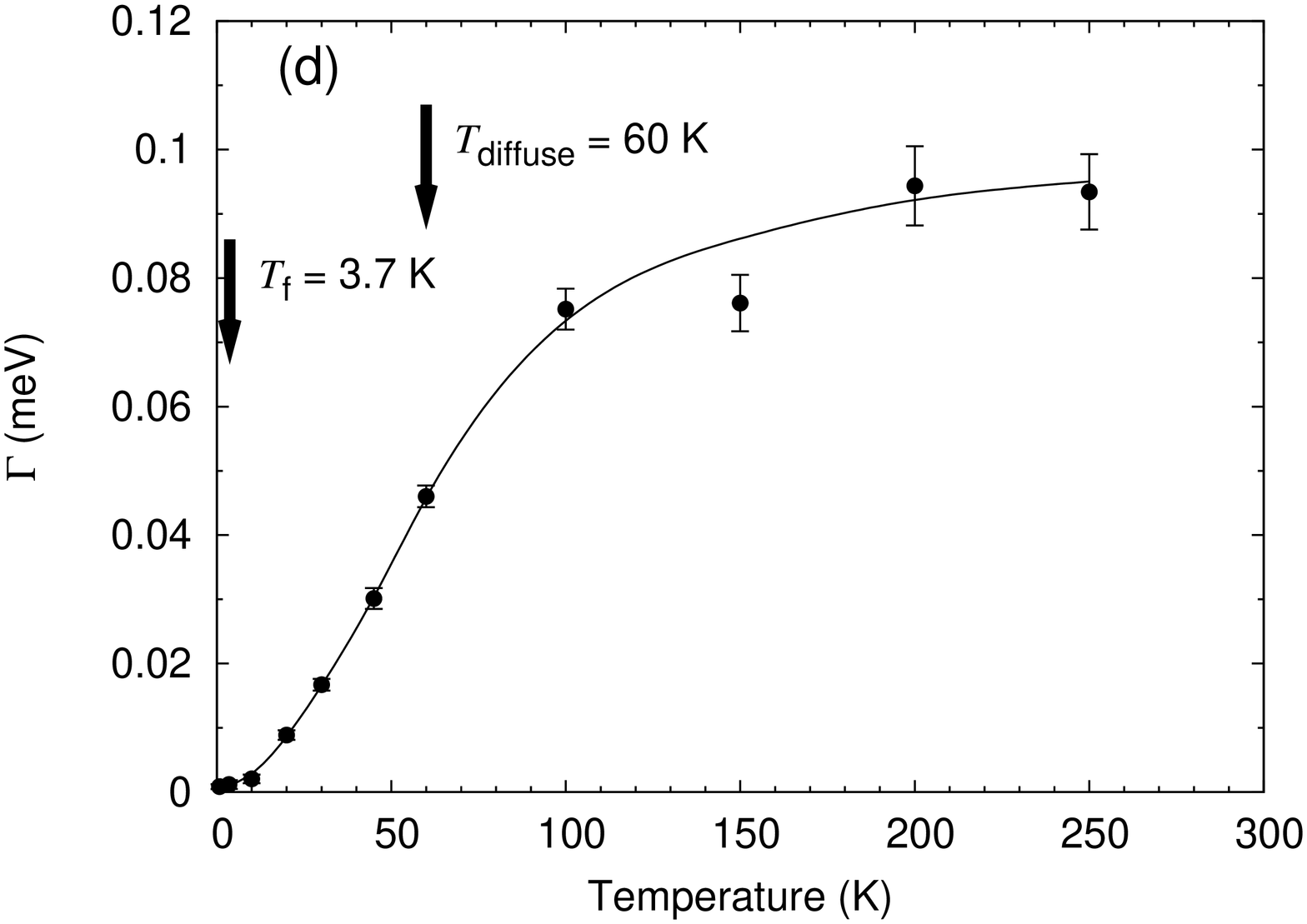}
\caption{(a) Temperature dependence of diffuse scattering intensity observed with higher energy resolution $\Delta E = 0.1$~meV.
Inset: Temperature dependence of diffuse scattering intensity at $2\theta = 20.3^{\circ}$ of Zn-Mg-Tb quasicrystal measured at LAM-40 with energy resolution $\Delta E = 0.32$~meV. The solid lines are guides to the eye.
(b) Temperature dependence of quasielastic spectra between 250~K and 0.7~K.
(c) Some fitting results of the quasielastic spectra showing contributions of total, elastic, quasielastic and constant background functions represented with solid, dash-dotted, dashed and dotted lines.
In (a), (b) and (c), scattering intensity is integrated in $Q$-range of $0.74 < Q < 0.96$~\AA$^{-1}$ by horizontally focusing analyzer.
(d) Temperature dependence of quasielastic peak width ${\it \Gamma}$ (half-width at half-maximum) obtained by fitting the spectra of (b) to a model scattering function (\ref{eq:modelSw}).
}
\label{C11inelneutron}
\end{figure}

\begin{figure}[htbp]
\includegraphics[angle=-90, width=60mm]{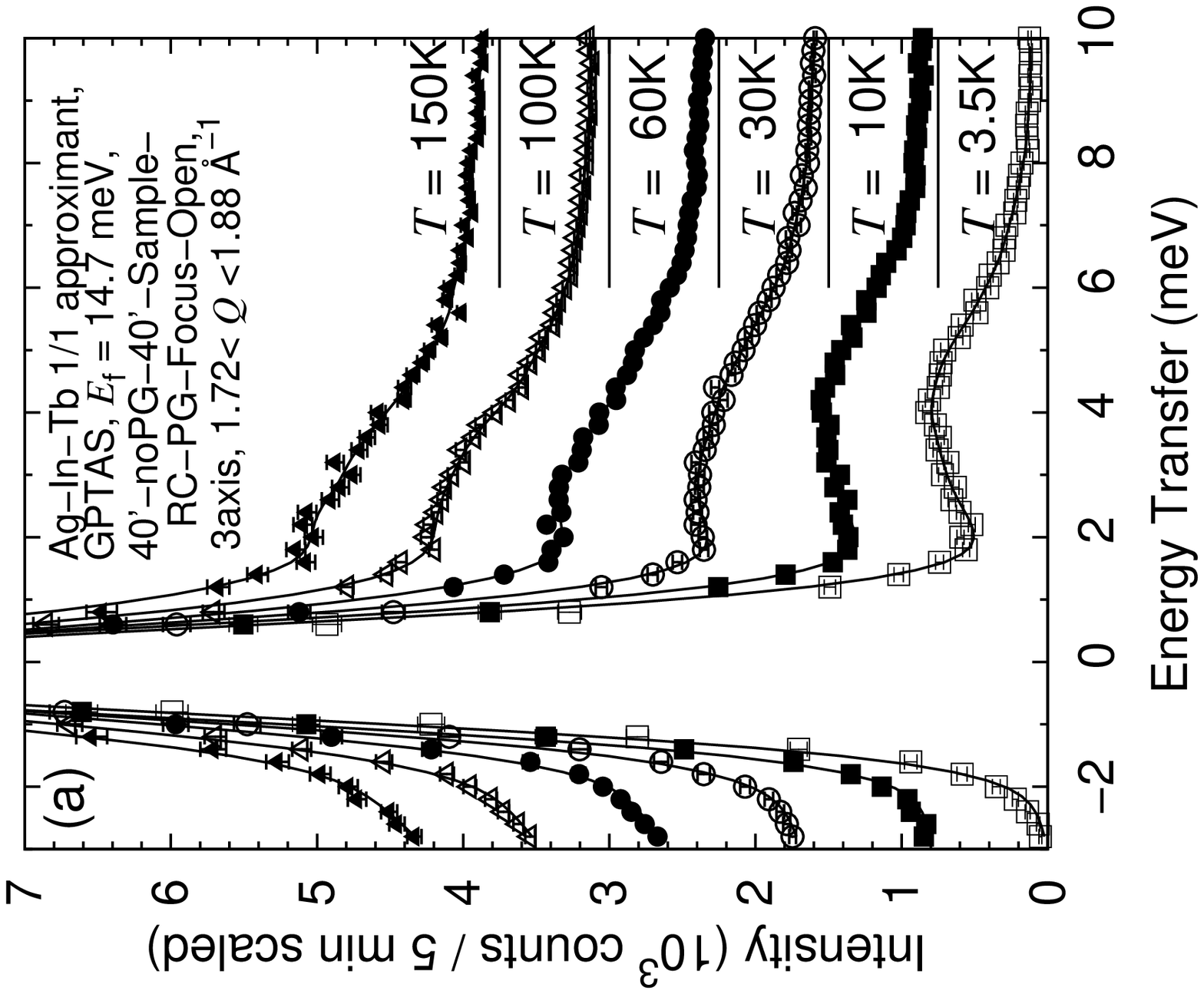}
\includegraphics[angle=-90, width=80mm]{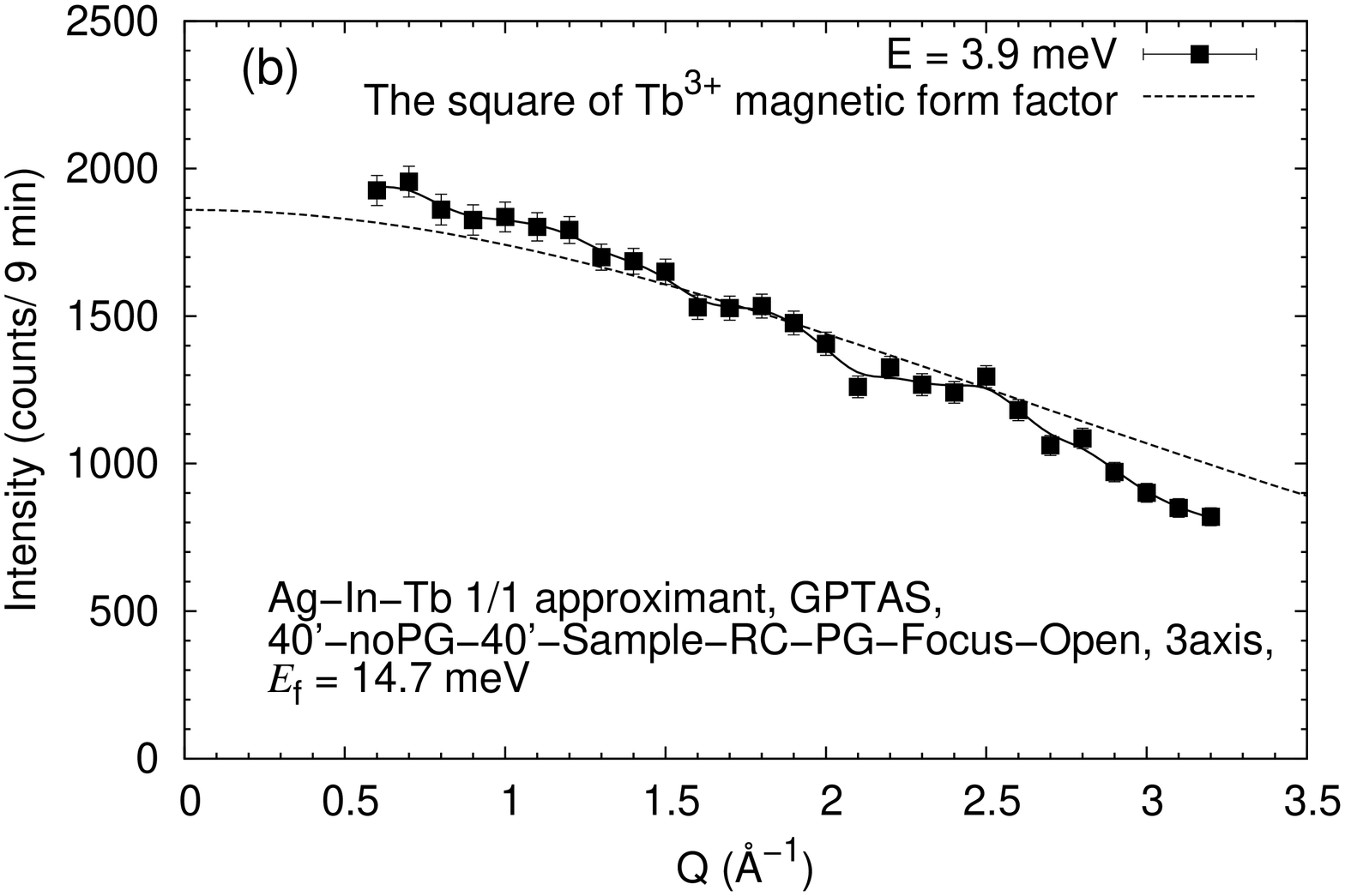}
\caption{(a) Neutron inelastic scattering spectra at several temperatures between 150~K and 3.5~K with coarse energy resolution $\Delta E = 1.3$~meV.
Scattering intensity is integrated in $Q$-range of $1.72 < Q < 1.88$~\AA$^{-1}$ by horizontally focusing analyzer. 
Solid lines are guides to the eye.
(b) $Q$-dependence of broad inelastic peak at $T = 3.5$~K and $\hbar\omega = 3.9$~meV.
The solid line is a guide to the eye and the dashed line stands for the square of Tb$^{3+}$ magnetic form factor.
}
\label{4Ginelneutron}
\end{figure}

To investigate the spin freezing behaviour at low temperature microscopically, we have performed neutron elastic and inelastic scattering experiments.
For these neutron experiments, we selected RE = Tb sample because of the highest freezing temperature. This means that energy scale is the highest among Ag-In-RE 1/1 approximants and makes inelastic experiment feasible.

 Neutron powder diffraction patterns observed at the lowest temperature $T = 3$~K and paramagnetic temperature $T = 60$~K are shown in figure~3(a). 
At 60~K, all the Bragg peaks were indexed by nuclear reflections of the 1/1 approximant phase, as shown by vertical solid lines in the bottom part of the figure.
It follows that impurity phase in the sample used in the neutron experiment is too small to detect obviously.
On cooling to 3~K, no additional Bragg reflection appears in the diffractogram.
This confirms the absence of long-range magnetic order in the Ag-In-Tb 1/1 approximant.
On the other hand, there is a weak but definitely observable increase of magnetic diffuse scattering intensity at the lowest temperature.
To visualize this increase more clearly, the temperature difference between the two diffraction patterns, or $I(3~{\rm K}) - I(60~{\rm K})$, are shown in figure~3(b).
Broad peaks are apparently seen at $Q = 0.85$ and $1.8$~\AA$^{-1}$.
The appearance of diffuse scattering peaks suggests development of magnetic short-range order at the base temperature.
The temperature difference becomes negative at $Q \rightarrow 0$, indicating that dominant spin correlations are antiferromagnetic, which is consistent with the negative Weiss temperature $\theta$.
From width of the first diffuse scattering peak at $Q = 0.85$~\AA$^{-1}$, we estimate the correlation length as $\xi \sim 9$~\AA(FWHM). 
It is noted that a diameter of a single icosahedral cluster of Tb ions in the Ag-In-Tb 1/1 approximant phase is approximately 11~\AA, and thus the observed spin correlation length corresponds to the diameter.

Next, evolution of the short-range spin correlations was investigated by observing temperature dependence of the diffuse scattering peak at $Q = 0.85$~\AA$^{-1}$.
Since the peak was considerably broad in the $Q$ space, we utilized horizontally focusing analyzer with triple-axis mode to gain higher counting statistics in limited beam time, which collect scattering intensity in $Q$ range of $0.74 < Q < 0.96$~\AA$^{-1}$.
An observed spectrum is shown in figure~4(a).
It can be seen in the figure that the intensity of the diffuse scattering increases gradually below $T = T_{\rm diffuse} = 60$~K.
$T_{\rm diffuse}$ was determined by using a cross-over between the two linear regimes.
Hence, we can conclude that short-range spin correlations develop at the temperature $T_{\rm diffuse}$, which is significantly larger than the freezing temperature $T_{\rm f} = 3.7$~K.
In principle, $T_{\rm f}$ and $T_{\rm diffuse}$ can be different, because magnetometry and neutron scattering measurements detect fluctuations of different time scales (the order of a few seconds for the former, whereas $10^{-12}$~s for the latter).
The ratio of the two temperature scales, $T_{\rm f}/T_{\rm diffuse} \simeq 0.06$, is surprisingly smaller than those of canonical spin glasses~\cite{mur78}.

To elucidate this slowing-down process of spin fluctuations, the neutron quasielastic signal was investigated.
Several inelastic scattering spectra observed at $Q = 0.85$~\AA$^{-1}$ with $T = 0.75, 3.5, 10, 20, 30, 45, 60, 100, 150, 200$ and 250~K are shown in figure~4(b).
Quasielastic tails were observed at high temperature, in addition to a central purely elastic scattering of structural origin.
The spectra were then fitted, convoluted with the instrumental resolution, to the following scattering function including a Lorentzian spectral weight function with the width ${\it \Gamma}$:
\begin{equation}\label{eq:modelSw}
S(\hbar \omega; T) = C_{\rm elastic}\delta(\hbar\omega) + C_{\rm qel}\frac{1}{1-\exp(-\hbar\omega/k_{\rm B}T)}\frac{{\it \Gamma} \hbar \omega}{(\hbar\omega)^2 + {\it \Gamma}^2} + C_{\rm bg},
\end{equation}
where the first, second and third terms represent elastic, quasielastic and constant background functions. In the fitting procedure, $C_{\rm elastic}$ was fixed to a value determined by the spectrum at $T=$250~K. Some fitting results showing each contribution are shown in figure~4(c). Solid, dash-dotted, dashed and dotted lines represent total, elastic, quasielastic and background functions.
Temperature dependence of the obtained half-width at half maximum ${\it \Gamma}$ is shown in figure~4(d).
In higher temperature range $100 < T < 250$~K, ${\it \Gamma}$ shifted moderately.
In contrast, the width suddenly decreased below 100~K. This indicates that the spins drastically slow down in this temperature range.
This is consistent with the increase of the elastic signal below $T_{\rm diffuse} = 60$~K shown in figure~4(a).
Since the quasielastic width eventually becomes almost zero at a temperature close to $T_{\rm f}$, it is pointed out that the corresponding relaxation is indeed responsible for the freezing behaviour observed in the macroscopic measurements.

To obtain further insight into faster spin dynamics, we performed inelastic neutron scattering experiments in wider energy range up to 10~meV.
The experiment was carried out using the energy resolution of $\Delta E = 1.3$~meV. 
The resulting temperature variation of the inelastic spectrum is shown in figure~5(a).
At high temperature, such as $T = 150$~K, the inelastic spectrum was again dominated by a broad quasielastic signal centred at $\hbar \omega = 0$.
As temperature decrease, an inelastic scattering peak emerged below 60~K with the peak energy shifting to 4~meV toward the base temperature.
In view of this considerably broader energy width at high temperature, origin of the quasielastic signal should be interpreted as different from that of the narrow quasielastic peak observed with the higher energy-resolution experiment shown in figure~4(b).
In addition to the measurement, $Q$ dependence of the inelastic peak at $\hbar \omega = 3.9$~meV is shown in figure~5(b).
$Q$ dependence of intensity was correspond to the square of magnetic form factor of Tb$^{3+}$ ion~\cite{Fre1979}, confirming magnetic origin of the inelastic peak.
Apparently a crystalline field splitting is ruled out from the possibility of origin of the peak in view of temperature dependence of the peak and good linearity of the inverse magnetic susceptibility even at low temperature for RE = Tb.
It is emphasized that the temperature at which the inelastic peak emerge roughly corresponds to $T_{\rm diffuse}$, suggesting a close relation between formation of the short-range correlations and the inelastic peak.

\section{Discussion}
In the present study, we found that some of Ag-In-RE 1/1 approximants exhibit macroscopic spin freezing behaviour in magnetic susceptibility measurements, which is similar to that of canonical spin glasses.
On the other hand, three distinct features from canonical spin glasses have been revealed in the neutron scattering results for RE = Tb sample.
First, significant short-range-spin correlations appear at the lowest temperature, contrasted to completely random freezing in canonical spin glasses.
Second, development of the diffuse scattering starts at fairly higher temperature $T_{\rm diffuse} = 60$~K than the macroscopic freezing temperature $T_{\rm f} = 3.7$~K.
Third, another considerably broader quasielastic component appears at high temperature $T > T_{\rm diffuse}$ and changes into broad inelastic peak at $\hbar \omega = 4$~meV as the short-range correlation is formed.

These totally unique results suggest following two-step freezing in the 1/1 approximant:
first of all, at $T>T_{\rm diffuse}$, individual spin fluctuates with no spatial correlation, giving rise to the considerably broader quasielastic signal.
As first step freezing, below $T_{\rm diffuse}$, spatial correlations of spin fluctuations starts to develop. 
Spins in each magnetic cluster fluctuate coherently, or intra-cluster spin correlations develop, because the length scale of the short-range spin correlation corresponds to that of the icosahedral spin cluster of Tb ions, which is a characteristic magnetic block in the 1/1 approximant phase.
Moreover the clusters formed by correlated spins probably have collective excitation modes and lead to the inelastic peak at 4~meV. 
An enhancement of the intra-cluster spin coherence with cooling would leads the inelastic intensity increase and rise the energy up to 4 meV from 0 meV.
As second step, with further cooling, fluctuations of the spin clusters becomes slower and slower as shown by the width ${\it \Gamma}$ in the narrower quasielastic spectrum, and eventually, cluster fluctuations freeze at $T_{\rm f}$ of the order of seconds.
Thus we have cleared up the difference of the freezing behavior in the Ag-In-Tb 1/1 approximant from one seen in the canonical spin glasses.

In this paragraph, contrastively, we point out great similarities between the present approximant and the Zn-Mg-Tb magnetic quasicrystal.
First similarity in both the systems is that a well localized isotropic Tb$^{3+}$ ions and the random spin-glass-like freezing, indicated by a Curie-Weiss behaviour in a wide temperature range and apparent bifurcations at low temperature in macroscopic susceptibility measurements.
Variations of characteristic parameters are described below;
freezing temperature $T_{\rm f}$ and a Weiss temperature $\theta$ are 3.7~K and -34~K for the Ag-In-Tb 1/1 approximant of the present study, whereas 5.8~K and -26.3~K for the Zn-Mg-Tb quasicrystal~\cite{fis99}.
A frustration parameter $|\theta / T_{\rm f}|$ is about two times larger in the 1/1 approximant than the quasicrystal. Thus the approximant would have higher degree of frustration.
This may be due to weaker intrinsic structural disorder expected for the crystalline approximant phase.
Next similarity is existence of short-range spin correlation. Especially, the correlation length is of the same degree as a diameter of magnetic atom cluster in each sample; The correlation length at the base temperature is about 20~\AA(FWHM) for the quasicrystal, while 9~\AA(FWHM) for the approximant. The diameter of the magnetic cluster are 15~\AA~\cite{ishimasa04} and 11~\AA~respectively.
The relations lead to the idea that clusters commonly form rigid spin objects approximately by the cluster at low temperature.
Furthermore, temperature scale at which the short-range-correlated spin fluctuations fall into elastic time window of the neutron experiment is mostly the same; the diffuse scattering intensity increases around 60~K in both the systems as shown by figure~4(a).
Therefore, temperature scales of the diffuse scattering are not scaled by freezing temperature $T_{\rm f}$, but rather by Weiss temperatures $\theta$ which are mostly the same in the two systems.
This is quite reasonable, since it is a Weiss temperature which provides a rough estimate of strength of inter-spin interactions.
It is further suggested that the inter-spin interactions are of the similar strength in the two systems.
The third point is the broad inelastic peak at low temperature, which is at 4~meV in the approximant and 2~meV in the quasicrystal. The energy difference possibly reflects the variation of the cluster.

As seen above, the magnetic freezing behavior of formation of short-range spin correlations around 60~K and its freezing at the low temperature $T_{\rm f}$, is quite in common in the two seemingly different samples, i.e. the approximant with periodically arranged magnetic cluster and the quasicrystal with quasiperiodically arranged magnetic cluster.
This indicates evidently that a key to explain the freezing behaviour without long-range magnetic order in the two systems is the local highly symmetric cluster; dodecahedral cluster for the Zn-Mg-Tb quasicrystal and icosahedral for the Ag-In-Tb 1/1 approximant.
Thus, we would suggest that long-range quasiperiodicity is not the origin of the freezing behaviour in the magnetic quasicrystal.

\section{Summary}
We have performed combined magnetic susceptibility and neutron scattering measurements on the Ag-In-RE 1/1 approximants.
In the magnetic susceptibility measurements, it has been shown that for most of the RE elements, Ag-In-RE approximants have well localized isotropic moments. Exceptionally for RE = Ce and Pr, crystalline field splitting or valence-fluctuation effects would hinder linear Curie-Weiss behaviour. For RE = Sm and Yb, they are in non-magnetic divalent states.
Especially for RE = Eu, Gd, Tb　and Dy, which contains isotropic localized moments, a bifurcation of the FC and ZFC magnetic susceptibility was observed at low temperature of 3~K, suggesting a spin-glass-like freezing of the spins.
In the neutron scattering measurements for RE = Tb, significant diffuse scattering below $T_{\rm diffuse} = 60$~K was detected, which accompanied by the evolution of the broad inelastic peak at $\hbar \omega = 4$~meV.
From these measurements, two-step freezing behaviour was revealed. Below 60~K, first intra-cluster antiferromagnetic spin correlations in icosahedral clusters develop. Upon further cooling to low temperature $T_{\rm f} = 3.7$~K, the cluster-spin fluctuations freeze without inter-cluster long-range magnetic ordering.
The freezing process is semi-quantitatively identical to that observed in the Zn-Mg-Tb quasicrystal, indicating that origin of the spin freezing behaviour is not quasiperiodicity, but high symmetry of their magnetic cluster.

\ack
The present authors thank Drs. A.~P.~Tsai and H.~Takakura for valuable comments and stimulating discussions.
This work is partly supported by the Grand-in-Aid for the basic research (C) from MEXT, Japan.

\section*{References}
\bibliography{zmt}

\providecommand{\newblock}{}
\begin{thebibliography}{10}
\expandafter\ifx\csname url\endcsname\relax
  \def\url#1{{\tt #1}}\fi
\expandafter\ifx\csname urlprefix\endcsname\relax\def\urlprefix{URL }\fi
\providecommand{\eprint}[2][]{\url{#2}}

\bibitem{sato05}
Sato T~J 2005 {\em Acta Crystallogr. Sec.\/} A {\bf 61} 39--50

\bibitem{ved04}
Vedmedenko E~Y, Grimm U and Wiesendanger R 2004 {\em Phys. Rev. Lett.\/} {\bf
  93} 076407

\bibitem{wes03}
Wessel S, Jagannathan A and Haas S 2003 {\em Phys. Rev. Lett.\/} {\bf 90}
  177205

\bibitem{luo93}
Luo Z, Zhang S, Tang Y and Zhao D 1993 {\em Scripta Metall. Mater.\/} {\bf 28}
  1513--8

\bibitem{nii94}
Niikura A, Tsai A~P, Inoue A and Matsumoto T 1994 {\em Philos. Mag. Lett.\/}
  {\bf 69} 351--5

\bibitem{hat95}
Hattori Y, Niikura A, Tsai A~P, Inoue A, Masumoto T, Fukamichi K, Aruga-Katori
  H and Goto T 1995 {\em J. Phys.: Condens. Matter\/} {\bf 7} 2313--20

\bibitem{fis99}
Fisher I~R, Cheon K~O, Panchula A~F, Canfield P~C, Chernikov M, Ott H~R and
  Dennis K 1999 {\em Phys. Rev.\/} B {\bf 59} 308--21

\bibitem{sato06}
Sato T~J, Takakura H, Tsai A~P and Shibata K 2006 {\em Phys. Rev.\/} B {\bf 73}
  054417

\bibitem{mur81}
Murani A~P 1981 {\em J.\ Magn.\ Magn.\ Mater.\/} {\bf 22} 271--81

\bibitem{Guo2002}
Guo J~Q and Tsai A~P 2002 {\em Phil. Mag. Lett.\/} {\bf 82} 349--52

\bibitem{Rua04}
Ruan J~F, Kuo K~H, Guo J~Q and Tsai A~P 2004 {\em J. Alloys Comp.\/} {\bf 370}
  L23--7

\bibitem{mor08}
Morita Y and Tsai A~P 2008 {\em Jpn. J. Appl. Phys.\/} {\bf 47} 7975--9

\bibitem{Joh1964}
Johnson I 1964 {\em Trans. Metall. Soc. AIME\/} {\bf 230} 1485--7

\bibitem{Pal1971}
Palenzona A 1971 {\em J. Less-Common Met.\/} {\bf 25} 367--72

\bibitem{Tak2001}
Takakura H, Guo J~Q and Tsai A~P 2001 {\em Phil. Mag. Lett.\/} {\bf 81} 411--8

\bibitem{Gom2003}
Gomez C~P and Lidin S 2003 {\em Phys. Rev.\/} B {\bf 68} 024203

\bibitem{Guo2000}
Guo J~Q, Abe E and Tsai A~P 2000 {\em Phys. Rev.\/} B {\bf 62} R14605--8

\bibitem{Tsa2000}
Tsai A~P, Guo J~Q, Abe E, Takakura H and Sato T~J 2000 {\em Nature\/} {\bf 408}
  537--8

\bibitem{takakura07}
Takakura H, Gomez C~P, Yamamoto A, \mbox{de Boissieu} M and Tsai A~P 2007 {\em
  Nature Mater.\/} {\bf 6} 58--63

\bibitem{ishimasa98}
Ohno T and Ishimasa T 1998 {\em Proc. 6th Int. Conf. on Quasicrystals
  (Tokyo)\/} ed Takeuchi S and Fujiwara T (Singapore: World Scientific) p~39

\bibitem{ishimasa04}
Ishimasa T, Oyamada K, Arichika Y, Nishibori E, Takata M, Sakata M and Kato K
  2004 {\em J. Non-Cryst. Solids\/} {\bf 334\&335} 167--72

\bibitem{lam40}
Inoue K, Ishikawa Y, Watanabe N, Kaji K, Kiyanagi Y, Iwasa H and Kohgi M 1985
  {\em Nucl. Instrum. Methods Phys. Res.\/} A {\bf 238} 401--10

\bibitem{Var76}
Varma C~M 1976 {\em Rev. Mod. Phys.\/} {\bf 48} 219--38

\bibitem{mur78}
Murani A~P and Heidemann A 1978 {\em Phys. Rev. Lett.\/} {\bf 41} 1402--6

\bibitem{Fre1979}
Freeman A~J and Desclaux J~P 1979 {\em J. Magn. Magn. Mater.\/} {\bf 12} 11--21

\end{thebibliography}

\end{document}